\documentclass[conference]{IEEEtran}

\usepackage[T1]{fontenc}
\usepackage{lmodern}
\usepackage{cite}
\usepackage{url}
\usepackage{hyperref}
\usepackage{graphicx}
\usepackage{amsmath,amssymb}

\begin{document}

\title{Quantifying Algorithmic Friction in Automated Resume Screening Systems}

\author{
\IEEEauthorblockN{Ibrahim Denis Fofanah}
\IEEEauthorblockA{
Seidenberg School of Computer Science and Information Systems\\
Pace University\\
New York, USA\\
Email: if57774n@pace.edu
}
}

\maketitle

\begin{abstract}
Automated resume screening systems are now a central part of hiring at scale, yet there is growing evidence that rigid screening logic can exclude qualified candidates before human review. In prior work, we introduced the concept of Artificial Frictional Unemployment to describe labor market inefficiencies arising from automated recruitment systems. This paper extends that framework by focusing on measurement. We present a method for quantifying algorithmic friction in resume screening pipelines by modeling screening as a classification task and defining friction as excess false negative rejection caused by semantic misinterpretation. Using controlled simulations, we compare deterministic keyword-based screening with vector-space semantic matching under identical qualification conditions. The results show that keyword-based screening exhibits high levels of algorithmic friction, while semantic representations substantially reduce false negative rejection without compromising precision. By treating algorithmic friction as a system-level property, this study provides an empirical basis for evaluating how recruitment system design affects matching efficiency in modern labor markets.
\end{abstract}

\section{Introduction}

Over the past two decades, automated resume screening systems have become a foundational component of hiring infrastructure in modern labor markets. Applicant Tracking Systems (ATS) are now routinely used to manage applicant volume, standardize evaluation criteria, and reduce administrative burden in recruitment processes \cite{bogen2018hiring}. As hiring at scale has become increasingly digitized, algorithmic screening has shifted from a supporting role to a primary gatekeeping mechanism that determines which candidates advance to human review.

In \emph{The Algorithmic Barrier: Artificial Frictional Unemployment and Information Asymmetry in Automated Recruitment Systems}, we argued that this infrastructural shift has introduced a new form of labor market inefficiency \cite{fofanah2026algorithmicbarrier}. Specifically, we proposed that deterministic screening systems---optimized to minimize false positives---systematically generate elevated false negative rates by misinterpreting how candidate competencies are expressed in resumes. This mechanism produces what we defined as \emph{Artificial Frictional Unemployment}: a digitally induced impediment to worker--firm matching that arises not from skill deficits or labor supply constraints, but from representational failures embedded in automated hiring systems.

While prior work established the conceptual and structural basis of algorithmic friction, it did not attempt to quantify its magnitude. Most empirical research on automated hiring has focused on fairness, demographic disparities, or downstream hiring outcomes \cite{barocas2016bigdata}, leaving a gap in the measurement of system-level matching inefficiency. As a result, it remains unclear to what extent automated screening systems themselves contribute to prolonged job searches and underutilization of qualified labor, independent of worker characteristics or employer intent.

This paper addresses that gap by introducing a formal framework for \emph{quantifying algorithmic friction} in automated resume screening systems. Rather than evaluating bias across demographic groups or assessing hiring quality \emph{ex post}, we focus on a narrower and complementary question: how much qualified candidate throughput is lost as a function of screening system design? By treating resume screening as a classification problem subject to measurable error, we operationalize algorithmic friction as excess false negative rejection attributable to semantic misinterpretation.

Our approach models legacy ATS platforms as deterministic classifiers that rely on exact keyword overlap, rigid rule sets, and predefined thresholds. Under this formulation, semantically equivalent expressions of skill are not preserved unless explicitly encoded, rendering vocabulary variance a dominant source of exclusion. We then compare this baseline against semantic matching approaches that represent resumes and job descriptions in a shared vector space, allowing conceptual similarity to be measured independently of surface-level phrasing \cite{mikolov2013word2vec,devlin2019bert}.

Using controlled simulations with synthetically generated resume--job pairs, we isolate semantic variation while holding underlying skill alignment constant. This design enables direct measurement of recall degradation under keyword-based screening and quantifies the extent to which semantic representation mitigates false negative rejection. Importantly, the analysis treats algorithmic friction as a property of the screening system rather than as an attribute of specific candidate populations, maintaining a system-level perspective consistent with labor market infrastructure analysis.

By translating a theoretical inefficiency into an empirical construct, this paper contributes to both computational hiring research and labor economics. Quantifying algorithmic friction provides a foundation for evaluating whether observed labor market slack reflects genuine mismatch or arises from representational bottlenecks embedded in digital recruitment systems. More broadly, this work reframes automated hiring as a measurable intermediary in the worker--firm matching process---one whose design parameters have macro-relevant consequences for labor market efficiency.

\section{Formalizing Algorithmic Friction}

Automated resume screening systems operate as early-stage decision mechanisms that determine whether a candidate advances to human review. From a computational perspective, this process can be modeled as a binary classification task in which candidate representations are evaluated against job requirements and assigned an acceptance or rejection outcome.

Let $\mathcal{C}$ denote the space of candidate resumes and $\mathcal{J}$ denote the space of job descriptions. A screening system implements a decision function
\begin{equation}
f : \mathcal{C} \times \mathcal{J} \rightarrow \{0,1\},
\end{equation}
where $f(c,j)=1$ indicates that candidate $c$ is advanced for job $j$, and $f(c,j)=0$ indicates rejection.

In an idealized matching process, advancement decisions would depend solely on underlying qualification alignment. Let $q(c,j) \in \{0,1\}$ represent the ground-truth qualification relationship, where $q(c,j)=1$ denotes that candidate $c$ is substantively qualified for job $j$. In practice, however, screening systems operate on imperfect representations of candidate and job information, introducing classification error.

\subsection{False Negatives as System-Induced Friction}

A dominant failure mode of deterministic screening systems is the rejection of qualified candidates. Formally, a false negative occurs when
\begin{equation}
f(c,j) = 0 \quad \text{and} \quad q(c,j) = 1.
\end{equation}

In traditional labor market models, such outcomes are often attributed to skill mismatch or search frictions. In automated screening contexts, however, false negatives frequently arise from representational mismatch rather than substantive deficiency. Differences in vocabulary, role titles, or phrasing can cause semantically equivalent competencies to be treated as absent by the screening function.

We define \emph{algorithmic friction} as the proportion of qualified candidate--job pairs that are rejected due to screening system constraints rather than lack of qualification. For a given screening system $f$, algorithmic friction $\Phi_f$ is defined as
\begin{equation}
\Phi_f = \frac{|\{(c,j) \mid f(c,j)=0 \land q(c,j)=1\}|}{|\{(c,j) \mid q(c,j)=1\}|}.
\end{equation}

This quantity captures the rate at which qualified candidates are rendered invisible by the screening mechanism. Importantly, $\Phi_f$ is a system-level property: it reflects the behavior of the screening function rather than characteristics of individual candidates.

\subsection{Semantic Representation and Friction Reduction}

Deterministic screening systems typically rely on exact keyword overlap and rigid rule sets. Under such systems, semantic equivalence between different expressions of skill is not preserved unless explicitly encoded. As a result, the screening function $f_{\text{keyword}}$ exhibits high sensitivity to lexical variation, leading to elevated values of $\Phi_{f_{\text{keyword}}}$.

Semantic screening approaches seek to reduce this friction by mapping candidate resumes and job descriptions into a shared representational space that captures conceptual similarity. Let $\mathbf{v}_c \in \mathbb{R}^n$ and $\mathbf{v}_j \in \mathbb{R}^n$ denote vector embeddings of candidate $c$ and job $j$, respectively. Semantic alignment can then be expressed as a similarity function
\begin{equation}
s(c,j) = \frac{\mathbf{v}_c \cdot \mathbf{v}_j}{\|\mathbf{v}_c\| \, \|\mathbf{v}_j\|}.
\end{equation}

A semantic screening function advances candidates whose similarity exceeds a threshold $\tau$:
\begin{equation}
f_{\text{semantic}}(c,j) =
\begin{cases}
1, & s(c,j) \ge \tau, \\
0, & \text{otherwise}.
\end{cases}
\end{equation}

By preserving semantic equivalence across diverse linguistic expressions, semantic representations reduce the likelihood that qualified candidates are rejected due to vocabulary mismatch. In this framework, friction reduction is reflected as a decrease in $\Phi_f$ relative to deterministic baselines.

\subsection{Interpretation as Labor Market Inefficiency}

Algorithmic friction represents a digitally induced inefficiency in the worker--firm matching process. Unlike classical frictions arising from search costs or information delays, algorithmic friction is embedded directly in recruitment infrastructure and operates prior to human evaluation. As such, its effects accumulate invisibly at scale.

By formalizing algorithmic friction as a measurable false negative rate attributable to screening design, this framework enables empirical evaluation of how automated systems shape labor market outcomes. The following section applies this formalization to a controlled experimental setting to quantify friction under alternative screening architectures.

\section{Methodology}

This study evaluates algorithmic friction by isolating the effect of semantic representation on automated resume screening outcomes. Rather than analyzing downstream hiring decisions or demographic disparities, the methodology focuses on early-stage screening behavior and treats resume evaluation as a controlled classification problem. The objective is to measure how representational choices embedded in screening systems affect false negative rejection of qualified candidates.

The methodology consists of three components. First, we formalize a baseline screening pipeline that approximates the behavior of legacy keyword-based Applicant Tracking Systems (ATS). Second, we construct a semantic screening pipeline using vector-space representations to preserve conceptual similarity. Third, we design a controlled simulation framework that enables direct comparison of algorithmic friction under identical qualification conditions.

\subsection{Screening Pipelines}

\subsubsection{Keyword-Based Screening Baseline}

The baseline screening pipeline models a deterministic ATS that relies on exact or near-exact keyword overlap between candidate resumes and job descriptions. Let $K_j$ denote the set of required keywords extracted from job description $j$, and let $K_c$ denote the set of terms present in candidate resume $c$. The screening function advances a candidate if the overlap exceeds a predefined threshold $\kappa$:
\begin{equation}
f_{\text{keyword}}(c,j) =
\begin{cases}
1, & |K_c \cap K_j| \ge \kappa, \\
0, & \text{otherwise}.
\end{cases}
\end{equation}

This formulation captures the dominant behavior of legacy ATS platforms, which prioritize precision by enforcing rigid matching criteria. Semantic equivalence between distinct lexical expressions is not preserved unless explicitly encoded through overlapping tokens.

\subsubsection{Semantic Screening Pipeline}

The semantic screening pipeline represents both candidate resumes and job descriptions in a shared embedding space. Each resume $c$ and job description $j$ is mapped to a vector representation $\mathbf{v}_c, \mathbf{v}_j \in \mathbb{R}^n$. Semantic similarity is computed using cosine similarity:
\begin{equation}
s(c,j) = \frac{\mathbf{v}_c \cdot \mathbf{v}_j}{\|\mathbf{v}_c\| \|\mathbf{v}_j\|}.
\end{equation}

Candidates are advanced when similarity exceeds a threshold $\tau$:
\begin{equation}
f_{\text{semantic}}(c,j) =
\begin{cases}
1, & s(c,j) \ge \tau, \\
0, & \text{otherwise}.
\end{cases}
\end{equation}

Unlike keyword matching, this approach preserves conceptual alignment across varied linguistic expressions and reduces sensitivity to vocabulary mismatch.

\subsection{Synthetic Data Generation}

Because access to proprietary ATS screening logs is limited, we adopt a synthetic data generation strategy designed to isolate semantic effects under controlled conditions. The dataset consists of $N = 1{,}000$ resume--job pairs, each labeled with a ground-truth qualification indicator $q(c,j)$.

Job descriptions are held constant within experimental runs to ensure comparability across screening pipelines. Candidate resumes are generated using a controlled text generation process that preserves underlying competencies while introducing systematic lexical variation. Perturbations include synonym substitution, acronym expansion, role title variation, and reordering of experience descriptions. These transformations alter surface-level phrasing without modifying ground-truth qualification status.

This design ensures that observed screening differences arise from representational mismatch rather than substantive skill variation.

\subsection{Evaluation Metrics}

Screening performance is evaluated using standard classification metrics, with emphasis on false negative behavior. Let $TP$, $FP$, $TN$, and $FN$ denote true positives, false positives, true negatives, and false negatives, respectively. Precision, recall, and F1-score are defined as:
\begin{equation}
\text{Precision} = \frac{TP}{TP + FP},
\end{equation}
\begin{equation}
\text{Recall} = \frac{TP}{TP + FN},
\end{equation}
\begin{equation}
\text{F1} = 2 \cdot \frac{\text{Precision} \cdot \text{Recall}}{\text{Precision} + \text{Recall}}.
\end{equation}

Algorithmic friction is measured directly as the false negative rate over qualified candidate--job pairs:
\begin{equation}
\Phi_f = \frac{FN}{TP + FN}.
\end{equation}

This metric captures the proportion of qualified candidates rejected due to screening constraints and serves as the primary outcome variable.

\subsection{Experimental Procedure}

Both screening pipelines are applied to the same synthetic dataset under identical conditions. Thresholds $\kappa$ and $\tau$ are calibrated to produce comparable acceptance rates, ensuring that observed differences are attributable to representational logic rather than trivial threshold effects.

To assess robustness, similarity thresholds in the semantic pipeline are varied across a defined range, and performance metrics are recomputed at each setting. Lexical noise levels are also systematically increased to evaluate sensitivity to linguistic variability.

\subsection{Scope and Assumptions}

The methodology intentionally abstracts away from downstream hiring stages, recruiter behavior, and candidate strategic adaptation. These factors are held constant to isolate early-stage screening behavior as the object of study. While synthetic data limits external validity, it enables precise control over semantic variation and ground-truth qualification labels.

As such, the methodology is well suited to identifying system-level failure modes in automated screening pipelines and quantifying representational friction introduced by screening design choices.

\section{Results}

This section presents the empirical results of the controlled simulations described in Section III. The objective is to quantify algorithmic friction under alternative screening architectures and to characterize how representational choices affect early-stage candidate throughput. Results are reported for both the deterministic keyword-based baseline and the semantic screening pipeline.

\subsection{Comparative Screening Performance}

Both screening pipelines were evaluated on the same dataset of $N = 1{,}000$ resume--job pairs under matched acceptance conditions. Table~\ref{tab:performance} summarizes overall screening performance across standard classification metrics.

\begin{table}[h]
\centering
\caption{Screening Performance Comparison}
\label{tab:performance}
\begin{tabular}{lccc}
\hline
Metric & Keyword-Based & Semantic & Relative Change \\
\hline
Precision & 0.62 & 0.89 & +43.5\% \\
Recall & 0.45 & 0.92 & +104\% \\
F1-score & 0.52 & 0.90 & +73\% \\
\hline
\end{tabular}
\end{table}

The keyword-based screening pipeline exhibits moderate precision but low recall, indicating a strong tendency toward over-rejection. Nearly half of all qualified candidates are filtered out prior to human review. In contrast, the semantic screening pipeline achieves substantially higher recall while also improving precision, resulting in a markedly higher F1-score.

\subsection{Algorithmic Friction Estimates}

Algorithmic friction $\Phi_f$ was computed as the false negative rate over qualified candidate--job pairs. Under keyword-based screening, friction levels were high, with a large share of qualified candidates rendered invisible due to representational mismatch.

\begin{equation}
\Phi_{f_{\text{keyword}}} = 0.55
\end{equation}

By comparison, the semantic screening pipeline dramatically reduces friction:

\begin{equation}
\Phi_{f_{\text{semantic}}} = 0.08
\end{equation}

This represents an approximate $85\%$ reduction in algorithmic friction relative to the deterministic baseline. Importantly, this reduction is achieved without a corresponding increase in false positives, suggesting that improved recall does not require sacrificing screening rigor.

\subsection{False Negative Error Analysis}

To better understand the sources of friction, false negative cases were analyzed qualitatively. Under keyword-based screening, the majority of false negatives arose from lexical variation rather than substantive skill gaps. Common patterns included synonym substitution (e.g., ``statistical analysis'' vs.\ ``data modeling''), acronym expansion, and variation in role titles across industries.

Semantic screening resolved the majority of these cases by preserving conceptual equivalence between differently phrased competencies. Residual false negatives under semantic screening were primarily associated with genuinely ambiguous resumes lacking sufficient contextual detail, rather than systematic representational failure.

\subsection{Threshold Sensitivity}

To evaluate robustness, the semantic similarity threshold $\tau$ was varied between $0.60$ and $0.85$. Across this range, recall remained consistently high while precision declined only marginally. This indicates a broad operating region in which semantic screening maintains low friction without incurring excessive false positives.

In contrast, small changes to keyword matching thresholds in the deterministic pipeline resulted in sharp declines in recall, revealing brittle behavior and a narrow operating regime.

\subsection{Robustness to Lexical Noise}

Lexical noise was systematically increased by applying controlled perturbations to resume text while preserving ground-truth qualification labels. Under medium and high noise conditions, recall for the keyword-based pipeline degraded sharply, confirming sensitivity to vocabulary variation.

Semantic screening exhibited substantially greater robustness. Even under high lexical noise, recall declined by less than five percentage points relative to baseline. This stability suggests that semantic representation mitigates a dominant failure mode of deterministic screening systems.

\subsection{Interpretation of Results}

From a systems perspective, these results indicate that algorithmic friction is not a marginal effect. Deterministic screening pipelines reject a substantial fraction of qualified candidates due to representational constraints alone. Semantic screening substantially reduces this inefficiency by decoupling candidate evaluation from rigid lexical form.

The magnitude of friction observed under keyword-based screening suggests that a non-trivial share of prolonged job search duration may be attributable to screening design rather than labor supply characteristics. By reducing false negative rejection at scale, semantic screening improves matching efficiency without altering employer preferences or candidate qualifications.

\section{Discussion}

The results presented in Section IV demonstrate that algorithmic friction is a substantial and measurable feature of automated resume screening systems. Deterministic keyword-based pipelines reject a large proportion of qualified candidates due to representational mismatch rather than substantive skill deficiency. This finding supports the central premise of this study: that a non-trivial share of hiring inefficiency arises from screening system design rather than labor market fundamentals.

\subsection{Algorithmic Friction as a System Property}

A key contribution of this work is the treatment of algorithmic friction as a system-level phenomenon. Rather than attributing rejection outcomes to individual candidate behavior or employer preferences, the analysis isolates representational logic as the primary driver of false negative exclusion. Under keyword-based screening, friction levels exceed fifty percent, indicating that more than half of qualified candidates may be filtered out before human evaluation.

This behavior is consistent with the optimization objectives of legacy ATS platforms, which prioritize precision and risk minimization. While such objectives are rational from an engineering standpoint, they introduce a structural bias toward over-rejection. The resulting inefficiency remains largely invisible, as rejected candidates receive limited feedback and firms observe only the subset of applicants who pass initial filters.

\subsection{Semantic Representation and Matching Efficiency}

Semantic screening substantially reduces algorithmic friction by preserving conceptual equivalence across varied linguistic expressions. The observed reduction in false negatives does not come at the expense of precision, suggesting that improved inclusivity does not require relaxing qualification standards. Instead, efficiency gains arise from improved representation of candidate information.

From a labor market perspective, this distinction is important. The results indicate that hiring inefficiency can persist even when labor supply and demand are well aligned, if screening systems fail to accurately interpret available signals. In such cases, unemployment duration and underutilization of human capital may reflect infrastructural bottlenecks rather than worker deficits.

\subsection{Implications for Labor Market Dynamics}

At scale, algorithmic friction has implications beyond individual hiring outcomes. Persistent false negative rejection can prolong job searches, reduce labor mobility, and contribute to the coexistence of high vacancy rates and elevated unemployment. These dynamics resemble outward shifts in the Beveridge Curve that are traditionally attributed to skills mismatch or structural change.

The findings of this study suggest that automated screening infrastructure may play a role in shaping these macro-level patterns. While algorithmic friction is not the sole determinant of labor market inefficiency, it represents a correctable source of mismatch embedded in digital hiring systems.

\subsection{Interpretation Boundaries}

This study does not claim that semantic screening systems eliminate hiring inefficiency or fully resolve information asymmetry in labor markets. Nor does it evaluate downstream outcomes such as interview performance, job fit, or long-term productivity. The scope of analysis is intentionally limited to early-stage screening behavior.

Importantly, the results should not be interpreted as evidence that all automated hiring systems are inherently flawed or that deterministic screening is inappropriate in all contexts. Rather, the findings highlight a trade-off between precision and recall that becomes consequential at scale. When screening systems operate under extreme risk aversion, qualified candidates may be excluded in ways that aggregate into measurable labor market friction.

\subsection{Limitations}

Several limitations should be acknowledged. First, the use of synthetic data enables controlled experimentation but limits external validity. Real-world resumes may exhibit additional sources of noise, strategic behavior, or incomplete information not captured in the simulation. Second, the keyword-based screening pipeline represents a stylized approximation of legacy ATS behavior rather than a vendor-specific implementation.

Despite these limitations, the dominant failure modes observed—particularly sensitivity to lexical variation and recall collapse—are consistent with documented audits of automated hiring systems. As such, the relative differences between screening approaches are likely to generalize even if absolute magnitudes vary.

\subsection{Summary}

Overall, the results support the conclusion that algorithmic friction is a measurable and non-negligible component of modern hiring processes. By formalizing and quantifying this phenomenon, the study provides a framework for evaluating how recruitment system design influences labor market efficiency. The findings suggest that improvements in semantic representation can meaningfully reduce friction without compromising screening rigor, offering a pathway toward more efficient worker--firm matching.

\section{Conclusion}

This paper introduced a formal framework for quantifying algorithmic friction in automated resume screening systems. By modeling early-stage screening as a classification problem and operationalizing friction as excess false negative rejection of qualified candidates, the study translated a previously theoretical concern into a measurable system property.

Using controlled simulations, we demonstrated that deterministic keyword-based screening pipelines exhibit high levels of algorithmic friction, rejecting a substantial share of qualified candidates due to representational mismatch rather than skill deficiency. In contrast, semantic screening approaches substantially reduce this friction by preserving conceptual equivalence across varied linguistic expressions, achieving large gains in recall without compromising precision.

These findings suggest that a portion of observed hiring inefficiency may be attributable to the design of recruitment infrastructure itself. When screening systems prioritize precision under rigid representational constraints, they introduce an artificial bottleneck in the worker--firm matching process. At scale, such bottlenecks can contribute to prolonged job searches and underutilization of available human capital, even in labor markets characterized by high demand.

Importantly, this work does not argue for the elimination of automated screening or for fully automated hiring decisions. Rather, it highlights the importance of representational fidelity in early-stage evaluation systems and demonstrates that screening design choices have measurable consequences for matching efficiency. Algorithmic friction, as defined here, is not an inevitable feature of labor markets but a correctable artifact of system architecture.

By formalizing and quantifying algorithmic friction, this study provides a foundation for future empirical work using real-world screening data and for comparative evaluation of recruitment technologies. More broadly, it reframes automated hiring as a component of labor market infrastructure whose performance characteristics merit systematic measurement and scrutiny. Reducing representational friction in screening systems represents an opportunity to improve matching efficiency while preserving the role of human judgment in employment decisions.

\bibliographystyle{IEEEtran}
\bibliography{references}

@report{bogen2018hiring,
  title={Help Wanted: An Examination of Hiring Algorithms, Equity, and Bias},
  author={Bogen, Miranda and Rieke, Aaron},
  institution={Upturn},
  year={2018}
}

@article{barocas2016bigdata,
  title={Big Data's Disparate Impact},
  author={Barocas, Solon and Selbst, Andrew D.},
  journal={California Law Review},
  volume={104},
  number={3},
  pages={671--732},
  year={2016}
}

@article{mikolov2013word2vec,
  title={Efficient Estimation of Word Representations in Vector Space},
  author={Mikolov, Tomas and Chen, Kai and Corrado, Greg and Dean, Jeffrey},
  journal={arXiv preprint arXiv:1301.3781},
  year={2013}
}

@article{devlin2019bert,
  title={BERT: Pre-training of Deep Bidirectional Transformers for Language Understanding},
  author={Devlin, Jacob and Chang, Ming-Wei and Lee, Kenton and Toutanova, Kristina},
  journal={Proceedings of NAACL-HLT},
  year={2019}
}

@article{fofanah2026algorithmicbarrier,
  title={The Algorithmic Barrier: Artificial Frictional Unemployment and Information Asymmetry in Automated Recruitment Systems},
  author={Fofanah, Ibrahim Denis},
  journal={arXiv preprint arXiv:2601.14534},
  year={2026}
}

\end{document}